\documentclass{aa}  

\usepackage{graphicx}
\usepackage{natbib}
\bibpunct{(}{)}{;}{a}{}{,}
\usepackage{txfonts}
\usepackage{gensymb}
\begin{document}

   \title{Testing a theoretical prediction for bar formation in galaxies with bulges}

\author{Sandeep Kumar Kataria 
        \inst {1,2} 
        \and
        Mousumi Das \inst{1}
        \and 
        Sudhanshu Barway \inst{1}
        }
        
    \institute{Indian Institute of Astrophysics, Koramangla, Bangalore-560034 \\
        \email{skkataria.iit@gmail.com}
               \and
               Indian Institute of Science, Bangalore 560012
              }
         
   \date{Received XX; accepted XX}
   \authorrunning {Kataria et al}

 
  \abstract{
  Earlier studies have shown that massive bulges impede bar formation in disk galaxies. Recent N-body simulations have derived a bar formation criterion that depends on the radial bulge force in a galaxy disk. We use those simulations to show that bars can form only when the force constant $FB < 0.13$, where $FB$ depends on the ratio of the bulge force to the total force of the galaxy at twice the disk scale length $2R_d$. In this article, we test this theoretical prediction using observational data obtained from the literature. Our sample consists of 63 barred galaxies with a wide range of Hubble classes from the $S^{4}G$ catalogue for which bulge, disk and bar decomposition has been done.  We find that 92 $\%$ of our sample galaxies satisfy the condition $FB < 0.13$ for bar formation in galaxies and hence agree with the bar formation criterion given by the simulations. }

   \keywords{Galaxies: statistics --
                Galaxies: spiral --
                (Galaxies:) bulges
               }

   \maketitle
%

\section{Introduction}

 Early studies have shown that a massive halo component can prevent the formation of bars in  disk galaxies \citep{OP.1973, Hohl.1976}. The presence of a massive halo makes a disk kinematically hot and prevents bar instabilities from forming.  It has also been shown that a massive central bulge component prevents bar formation as it introduces Inner Lindblad resonances \citep{Sellwood.1980} and cuts the feedback mechanism such that swing amplification doesn't work anymore \citep{Toomre.1981}. Apart from bulge mass, the effect of halo densities on bar formation have also found to be important; the corresponding criterion which depends on halo to disk mass ratios is known as the ENL criterion \citep{Efstathiou.1982}. However, the application of the ENL criteria for bar formation is limited because the criterion is based on only two dimensional disk simulations. Also the stellar velocity dispersion was not included and a rigid instead of a live dark matter halo was used to model the halo potential. These studies were limited by the availability of computational resources and  hence rigid halos were used. Later studies have also shown with various models that the  presence of a rigid halo does not allow the exchange of angular momentum between the disk and halo particles, which is important for the formation and evolution of bars \citep{Athanassoula.2008}.
 
 Other studies that have examined the effect of galaxy mass distributions on bars include the effect of the central mass concentration of galaxies on bars \citep{Norman.1996, Shen.2004,Athanassoula.2005a, Debattista.2006, Kataria.2019} which claim that bars become weaker with the increasing central mass concentration (CMC) until the mass is few percentage of the disk mass, at which point the bar dissolves. Massive compact CMCs \citep{Shen.2004} have been found to dissolve bars as they produce chaos in the large scale phase space region occupied by the bar while the lower mass CMCs make bars weak by producing chaos in the smaller phase space regions of the bar.
 
 Recent 3D N-body simulations of isolated disk galaxies with live dark matter halos have shown that both bulge mass and bulge concentration have a strong effect on bar formation timescales in disk galaxies \citep{KD.2018} (hereafter KD2018). The criterion depends on the ratio of the force due to the bulge and that due to the total galaxy, at the disk scale length $R_d$, and is given by $FB=\frac{GM_{bulge}}{{R_d}V^2_{tot}}$.  The study suggested that if FB$>$0.35, the disk does not form a bar when evolved under isolation.

Observational studies have also shown that there exists a correlation between bulges and bars, and a possible preference for bars to form in bulge dominated systems, although this maybe an observational bias as bulge dominated spiral galaxies are often easier to spot \citep{Sheth.2008, Skibba.2012}. In fact, the results of large optical surveys show that the bar fraction decreases with increasing bulge brightness \citep{Barazza.2008,Aguerri.2009}.  There is also a correlation between the central mass concentration within bulges in barred galaxies and bar ellipticity \citep{Das.2003,Das.2008}. Also, from observations of early to late type spiral galaxies along the Hubble Sequence, it is clear that bulges become less prominent  \citep{BT.2008}. Furthermore, some observations show that bulge to disk mass ratio decreases from early to late type spiral galaxies \citep{Laurikainen.2007,GW.2008}. The correlation of bars with bulge to disk flux ratios (B/D) is clearly seen in \cite{Laurikainen.2007} where the B/D ratio is smaller for early type barred spiral galaxies compared to early type non-barred spiral galaxies. Most of these studies indicate that bars should be less common in spirals with massive bulges, and their presence makes it harder to form bars in their disks. 

But the issue is more complicated than just bulge mass. For example studies such as  \citet{Diaz-Garcia.2016} show that bars are preferentially found in galaxies in which the mass is more centrally concentrated.  This suggests that bulge mass may not be the only important factor, but bulge mass concentration may also play an important role in the evolution of disk structure and bar formation in galaxy disks. 

Both observations \citep{Barazza.2008,Aguerri.2009} and simulations (KD2018) indicate that massive bulges do not allow bars to form easily in the disks of galaxies and that bulge concentration may also play a role. In this regard the bar formation criterion $FB = 0.35$, determined from simulations of disk galaxies by KD2018 is an important theoretical prediction that can be tested with observations of barred galaxies. The criterion is written in terms of parameters that can be easily obtained from observations  and includes the effect of both bulge mass and  concentration.

We also briefly discuss about the different types of bulges, their formation mechanism and properties since we have compared our numerical results with observations. There has been several discussions about the type of bulges in simulations compared to observations \citep{Athanassoula.2005,FD.2008,Erwin.2015}. \cite{Athanassoula.2005} has suggested that there are three types of bulges which are classical bulges, boxy/peanut-shaped bulges and disky pseudo bulges. Classical bulges are rounder objects and kinematically hotter, they are similar to elliptical galaxies and are thought to be formed by major mergers \citep{Kauffmann.1993,Baugh.1996,Hopkins.2009,Naab.2014}, multiple minor mergers \citep{Bournaud.2007,Hopkins.2010},accretion of smaller satellites \citep{Aguerri.2001} and monolithic collapse of a primordial cloud \citep{Eggen.1962}. Boxy/Peanut shape bulges are vertical thick systems and rotationally dominated in comparison to classical bulges. They are thought to be formed by disk instability during secular evolution \citep{Kormendy.2004}, vertical heating of bar due to buckling \citep{Combes.1990, Raha.1991, Martinez.2006} or heating of bar due to vertical resonances \citep{Pfenniger.1990}. Disky pseudo bulges are flattened system like an exponential disk in the nuclear region. They are thought to be formed by inward pulling of gas along the orbits and the consequent star formation \citep{Heller.1994,Regan.2004}. We have discussed the comparison between bulges of simulations (KD2018) and observations used in this article in section 3.

In this article, our aim is to obtain the value of $FB$ for a sample of barred galaxies in order to test the bar formation criterion mentioned above. For making a fair comparison, we have limited the sample of galaxies with bulge-to-total mass ratios similar to that of the simulated models in KD2018. The article is organized as follows. Section 2 summarizes the bar formation criterion. In Section 3, we describe the sample data, it's analysis and comparison of observational bulges to the simulated ones.  Section 4 is devoted to results and discussion. Summary of this work is presented in Section 5.


 \section{The Bar Formation Criterion - revised} 
\begin{figure}
    \centering
    \includegraphics[scale=0.45]{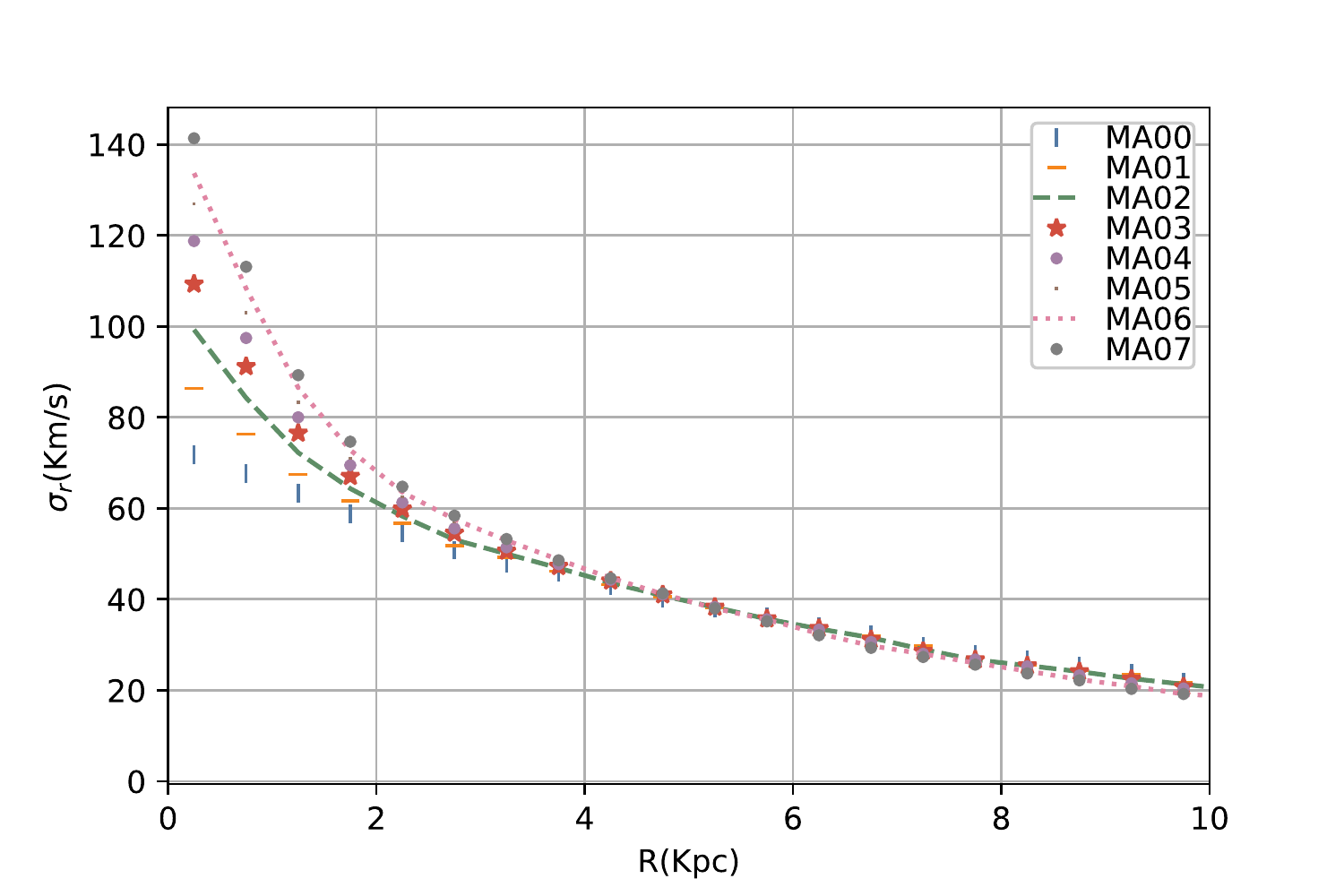}
    \caption{The radial velocity dispersion for all MA model with increasing bulge masses. Increasing bulge mass order is from MA00 to MA07; MA00 is without bulge and MB07 with highest bulge mass}
    \label{fig:MA}
\end{figure}

\begin{figure}
    \centering
    \includegraphics[scale=0.45]{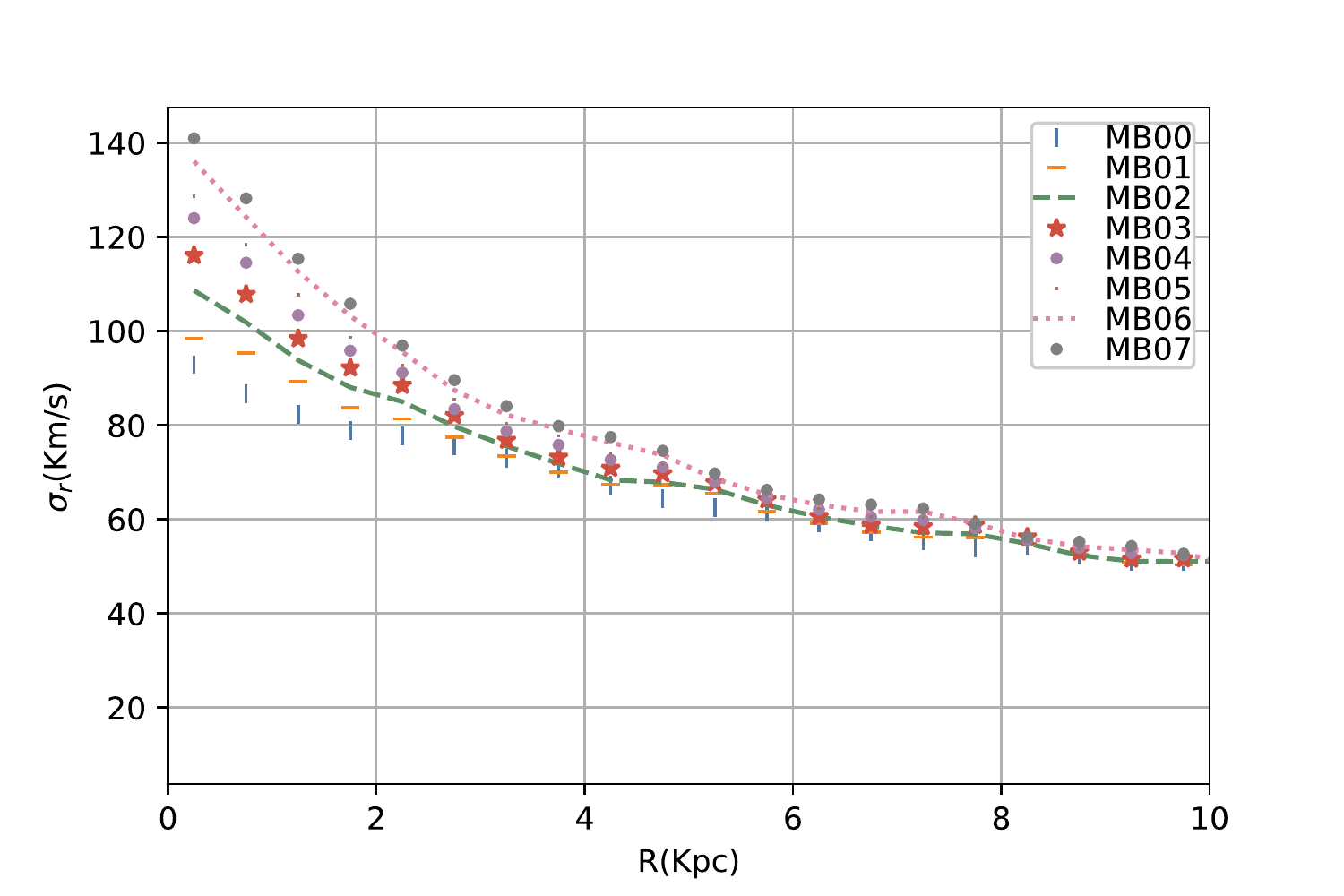}
    \caption{The radial velocity dispersion for all MB model with increasing bulge masses. Increasing bulge mass order is from MB00 to MB07; MB00 is without bulge and MB07 with highest bulge mass}
    \label{fig:MB}
\end{figure}

The simulations of KD2018 showed that both bulge mass and bulge concentration play an important role in inhibiting bar formation in disk galaxies. The bulge concentration is defined as the ratio of the effective bulge radius to disk scale length. The study included 2 types of disk models with increasing bulge masses called MA and MB, which differ in bulge concentrations. The disk scale lengths for models MB are almost twice that of models MA. Hence, MB models represent less concentrated galaxies compared to models MA (more details are given in KD2018). The study showed that as  bulge mass increased, bars could not form in the disks. This is  because a massive bulge makes a disk kinematically hotter as the central gravitational potential becomes deeper with increasing bulge mass. Figures \ref{fig:MA} and \ref{fig:MB} show this clearly; we find that as the bulge mass increases for MA and MB type models, it leads to a gradual increase in central velocity dispersion.

In KD2018, the models have bulge to disk mass ratios that vary from 0 to 0.7 or bulge to total galaxy mass ratios that vary from 0 to 0.41. The study showed that if the ratio of force due to the bulge component and that due to the total galaxy potential ($FB$) exceeds a value of 0.35, the bar instability will not develop in the disk. The relation is given by:
\begin{equation} 
     FB=\frac{F_b}{F_{tot}} = \frac{GM_{bulge}}{R_d V^2_{tot}}
     \label{FB0}
\end{equation}
 Here $M_{bulge}$ is the mass of the bulge component, $R_d$ is the disk scale length, $V_{tot}$ is the total rotational velocity at disk scale length. The study also showed that the bulge mass that inhibits bar formation is lower for models with dense bulges (MA) compared to models with less dense bulges (MB).
 
  However, KD2018 have used the initial values (i.e. before evolution) of disk scale length and rotation velocity to calculate the value of FB in the criterion. But the disk scale length and rotation curve change significantly as a disk forms a bar and secularly evolves till 9.79 Gyr. Hence, we have revised the calculations of FB using the final disk scale length and rotation curves to deduce FB values. In the revised version of the FB criterion we calculated the rotation velocity values at a radius which is twice that of the disk scale length. This is because this radius falls in the flat rotation curve of disk galaxies. We find that if FB exceeds a value of 0.13, the disk does not support a bar. Therefore, the revised value of the criterion is given by:
 \begin{equation} 
     FB=\frac{F_b}{F_{tot}} = \frac{GM_{bulge}}{2R_d V^2_{tot}}<0.13
     \label{FB}
\end{equation}
 Here $M_{bulge}$ is the mass of bulge component, $R_d$ is the disk scale length, $V_{tot}$ is the total rotational velocity at twice the disk scale length. 
 
The above formula can be applied to observations fairly easily as long as the bulge luminosity, disk scale length and galaxy rotation velocities are known for a sample of galaxies. 

\section{Data} \label{DA}

The first two parameters in equation [\ref{FB}], $M_{bulge}$ and $R_d$, can be obtained from the two dimensional (2D) bulge disk decomposition of a galaxy. For this purpose, we have used the bulge-disk decomposition of barred galaxies  provided by \citet{Salo.2015}. These 2D  decomposition's were done using 3.6$\mu$m images from the Spitzer Survey of Stellar Structure in Galaxies ($S^{4}G$)  using the GALFIT software \citep{Peng.2002, Peng.2010}. The study of \cite{Salo.2015} provides the bulge, disk and bar parameters of a sample of 103 barred galaxies along with reliable estimates of their bulge Sersic index and the bulge-to-total light ratio (B/T). 

The other parameter in the bar formation criterion is the rotation velocity at twice the disk scale length, $V_{tot}$. To determine this parameter we have used the Hyperleda \footnote{http://leda.univ-lyon1.fr/} database  \citep{Paturel.2003,Makarov.2014}. This constraint further reduces our sample size to 94 because of the unavailability of rotation curves. For some of the individual sources we have used HI line widths at full width half maximum (FWHM), also called the W50 value in the literature to determine $V_{tot}$. It is a measure of the total width at half the peak value of the HI line and is approximately twice the disk flat rotation velocity in a galaxy. These W50 values have been further corrected for galaxy inclination angle taken from Hyperleda database to obtain final flat rotation velocities $V_{tot}$.  These rotation velocities correspond to the outer region of galaxies. However, the $FB$ calculation needs rotation velocities at twice the disk scale lengths and so this approximation may result in lower values of the $FB$ constant in some galaxies.  A discussion for this limitation is given in section \ref{R}.

\begin{figure}
\centering
\includegraphics[scale=0.48]{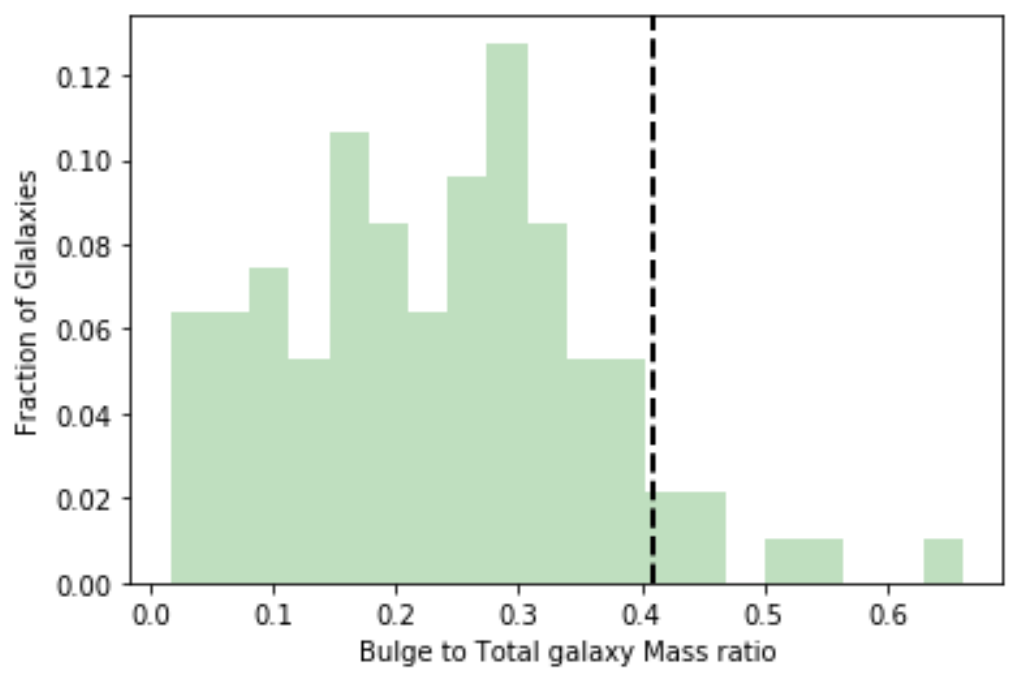}
\caption{The above figure shows the distribution of bulge to total disk mass for the sample galaxies. Here the vertical dashed line represents the upper limit of this ratio in the simulations of KD2018.}
\label{BT}
\end{figure}

In the simulations presented in KD2018, an upper limit of 0.41 for the bulge-to-total mass ratio was used. To obtain the bulge mass for our sample of galaxies, we have used the mass-to-light (M/L) ratio values of 0.5 \citep{Lelli.2016} to convert the bulge luminosity to bulge mass. The bulge luminosity is taken directly from the 2D decomposition of the galaxies \citep{Salo.2015} which derived the bulge luminosities as a fraction of the total galaxy luminosity. Figure \ref{BT} shows the distribution of our sample data with respect to bulge-to-total mass ratio. Here, the vertical dashed line denotes the upper limit of the bulge-to-total mass used in simulations (KD2018). In order to have a fair comparison, we have removed the galaxies which are outside the upper limit of bulge-to-total ratio used for simulated galaxies. This further reduces our sample to 87 galaxies. As nearly face on as well as edge on galaxies will lead to errors in de-projecting the rotation curves, we have only included galaxies that have inclination angles less than 80$^{\circ}$ and larger than 30$^{\circ}$. This restricts our final sample to 63 galaxies for conducting the study. Unless otherwise mentioned,  we have used WMAP cosmology \citep{Hinshaw.2013} for converting angular distances to physical distances.

 To compare our results of barred galaxies with an unbarred sample, we have also compiled the 2D decomposition parameters of an unbarred galaxy sample \citep{Salo.2015} to calculate FB criterion. We obtained 113 unbarred galaxies after putting similar constraints as we applied for the barred galaxies discussed above in this section i.e bulge-total mass fraction, inclination angle and availability of rotation curve etc.

 Data from \cite{Salo.2015} used for this study define bulges based on their Sersic indices and include both classical (n>2) and boxy/peanut bulges or pseudo bulges (n<2) \citep{FD.2008}. In the  KD2018 simulations, the galaxy models in the beginning of the simulations have bulges with hernquist profiles \citep{Yurin.2014} which are typically classical in nature.  As the galaxy models are evolved, the buckling of the bar introduces additional boxy bulges(B/P) \citep{Athanassoula.2005}. Therefore the KD2018 models contain composite bulges at the end \citep{Erwin.2015} which have both classical as well as boxy bulges. But the $FB$ criterion from KD2018 includes only the initial classical bulge mass determined from the beginning of simulations. The measured observational $FB$ criterion \citep{Salo.2015} may have a mixture of boxy/peanuts, disky pseudo bulges and classical bulges. Hence the observed $FB$ values will be an overestimate compared to the simulation ones.

\begin{figure}
\centering
\includegraphics[scale=0.48]{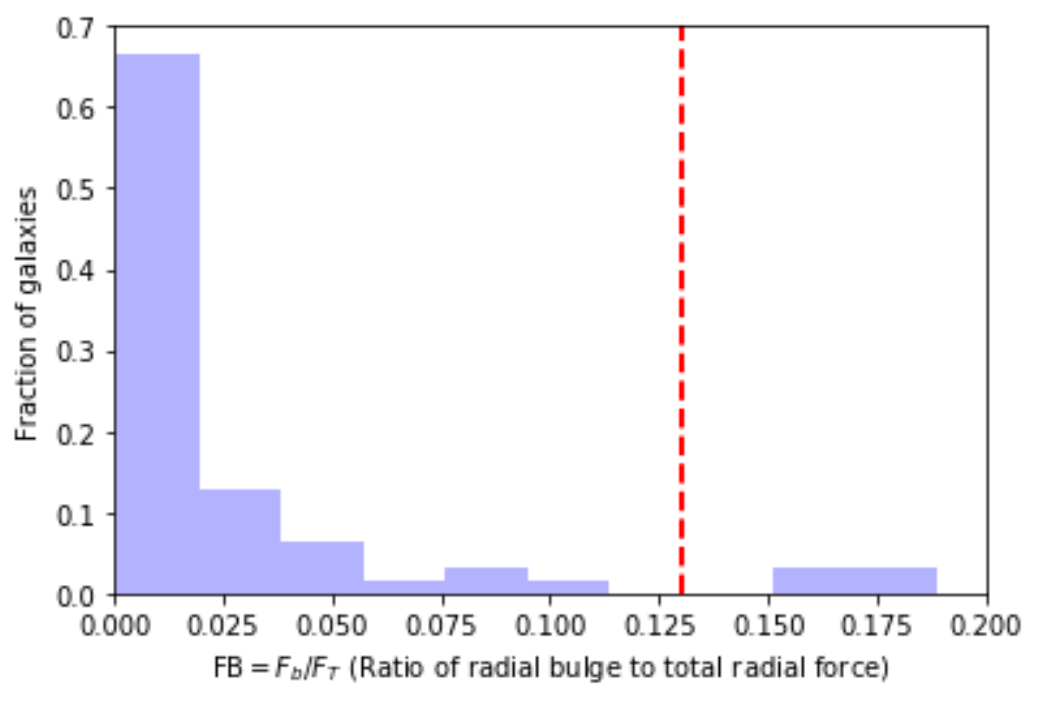}
\caption{Distribution of galaxies given according to the ratio FB i.e. ratio of radial bulge force to total radial force.}
\label{fig:FB1}
\end{figure}

\begin{figure}
    \centering
    \includegraphics[scale=0.48]{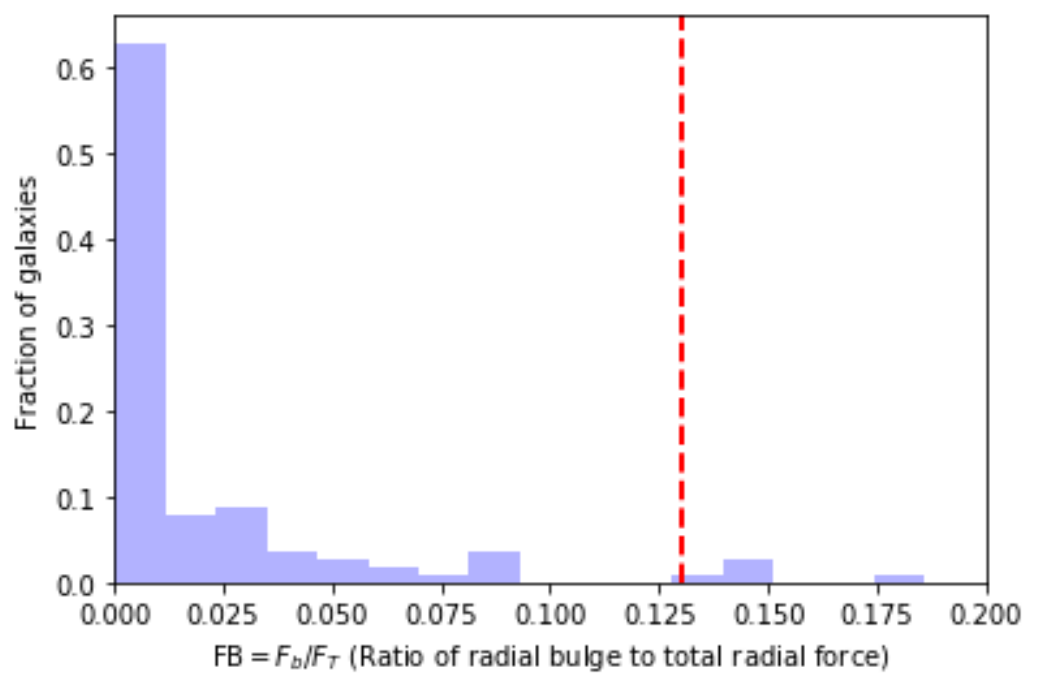}
    \caption{Distribution of unbarred galaxy according the ratio $FB$ parameter which is defined in equation \ref{FB}. } 
    \label{fig:unbarred}
\end{figure}

\section{Results} \label{R}
Once the observational data for the galaxy sample was obtained, we calculated the force constant $FB$ for our sample using the bulge masses, the disk scale lengths and the flat rotational velocities for each galaxy. In Fig. \ref{fig:FB1} we have plotted the histogram of the $FB$ parameter for the barred galaxy sample. We find that around 92 $\%$ of the galaxies in our sample lie below the value of 0.13 (Fig. \ref{fig:FB1}). This confirms that the value of $FB = 0.13$ represents an upper bound on the fractional radial bulge force which allows bar formation (KD2018). 

The barred galaxies with FB values larger than 0.13 are  ES0\,509-026, NGC\,3489, NGC\,4355, NGC\,6278. Of these ESO\,509-026 is interacting with the neighbouring galaxy IC\,4249 \citep{Telles.1995} and hence does not agree with the isolated evolution constraint of the bar formation criterion ($FB$ parameter). The galaxy NGC\,3489 is also part of the Leo group \citep{Watkins.2014} and is interacting with other galaxies within the group; hence it does not agree with the constraint of isolated galaxy evolution. For NGC\,4355, the HI detection maybe spurious \citep{Beckmann.2006} as the HI detection has been confused with the galaxy VV655 \citep{Thomas.2002}. The galaxy NGC\,6278 is in the phase of early stage merger activity \citep{Barrera-Ballesteros.2015} and hence again the isolated galaxy condition for $FB$ criteria is not satisfied for this galaxy.

In Fig. \ref{fig:unbarred} we have plotted the $FB$ criterion value for the unbarred galaxy sample. We find that surprisingly most of the unbarred galaxies are below the upper limit of bar formation, $FB = 0.13$. This is because the presence of a massive bulge is a necessary condition for inhibiting bar formation in galaxy disk but it is not a sufficient one. Therefore our bar formation criterion is a necessary condition for a disk to be stable but not a sufficient one as there are a few other factors  which are beyond the scope of the simulations (KD2018).

The main difference between the simulation models and the observations of galaxies is that in the $FB$ criterion models, the disk velocity dispersion is mainly determined by the bulge mass and the initial model parameters. But in real galaxies there are several possible methods by which a disk can increase it's velocity dispersion, such as : 1) tidal interactions with satellite galaxies \citep{Elmegreen.1995}, 2) when a galaxy passes through cluster potential \citep{Monica.1993}, 3) the accretion of a minor galaxy by a major galaxy which heats the stars in the major galaxy disk \citep{Qu.2011}. Also the 4) secular evolution of a bar - if a bar has dissolved and it is comparable to the bulge size, it will heat up the disk and bulge  \citep{Guo.2020}. Thus a disk maybe stable against bar formation despite the presence of an allowed bulge mass ($FB < 0.13$) as suggested by our criterion because of the higher velocity dispersion \citep{Athanassoula.2008}. This has an important implication for unbarred galaxies that satisfy $FB < 0.13$. It may mean that their disks are kinematically hot compared to similar galaxies that do not satisfy this criterion. 

\section{Discussion}
The HI line-widths $W_{50}$ used in this study gives an approximate estimate of two times the flat rotation velocity of the galaxy disk after correcting for inclination. However, the criterion given by equation [\ref{FB}] requires the rotation curve at disk scale length $2R_d$. In general $V_{tot}$ at $2R_d$ will be less than $\frac{1}{2}W_{50}$. Hence, our estimates of $FB$ are a lower limit to the values of equation [\ref{FB}]. For example, in the case of NGC\,4569, we initially used a flat rotation velocity $V_{tot}$ of 185 Kms$^{-1}$. However, the rotation curve shows that $V_{tot}$ at $2R_d$  corresponds to 179.6 Kms$^{-1}$ \citep{Sofue.1999}. Thus, in NGC\,4569 the $FB$ parameter changes from 0.022 to 0.024. However, both values satisfy the bar formation criterion ($FB < 0.13$). Similarly, if we assume that the rotation curves values at $2R_d$ are lower by 25$\%$ compared to the maximum value of the rotation curve, which is typically the case for most of the galaxies \citep{Sofue.1999}, our criterion will still be valid for 90 $\%$ of the sample galaxies.
 
It will be interesting to explore this bar formation criterion ($FB$) for our Galaxy, the Milky Way, which is a barred spiral galaxy \citep{Gerhard.2002}. Theoretical modelling of photometric and kinematic data \citep{BT.2008,McMillan.2011,Bovy.2013,Licquia.2016} predicts that the of disk scale length $R_d$ value of the Milky Way varies from 2 to 3 kpc. 
The rotation curve was obtained from \cite{Mcgaugh.2018}. There is an uncertainty about the presence of a classical bulge in our galaxy \citep{Bland-Hawthorn.2016}. We calculated the classical bulge mass which is required to satisfy $FB < 0.13$ at twice the disk scale length and found that it varies from 0.44 to 0.76 $ \times 10^{10} M_{\sun}$. This agrees with the upper limit for the allowed classical bulge mass in the Milky Way which is around 0.84 $ \times 10^{10} M_{\sun}$ \citep{Portail.2015} and certainly a much smaller mass in central region of Milky Way.  
 
The simulations of KD2018 show that the central mass concentration in a disk galaxy makes it kinematically hotter i.e. increases the velocity dispersion of disk stars. This can adversely affect bar formation as a bar will not form in a hot disk. We can see that barred galaxies obey the criterion defined by equation \ref{FB}. On the other hand we see that the unbarred galaxy sample does not seem to follow the bar formation criterion because it is a necessary condition for bar formation but is not a sufficient one. The reasons for the criterion not being sufficient are the factors other than the presence of bulge that affect disk velocity dispersion, such as tidal interaction, satellite merging etc. as discussed earlier, which make disks hotter.

The effect of the bulges on a disk not only depends on its mass, but also on the bulge concentration. In this study we haven't studied the observational effect of bulge concentration on our criterion because of different definitions of concentration in simulations(KD2018) and observations \citep{Barway.2016}. A more detailed observational study which will include wide sample containing barred and unbarred galaxies, is needed to test the effect of bulge mass and concentration in line with simulated results (KD2018). 

\section{Summary}\label{S} 
The primary motivation for this study was to test the bar formation criterion given by KD2018 with an observational data set. We have made a small revision to the threshold value of $FB$ for bar formation in a galaxy disk. We find bars form in galaxy disks only if the $FB < 0.13$ criterion is satisfied, where FB=$\frac{GM_{bulge}}{2R_d V^2_{tot}}$. We applied the criterion to a sample of 103 barred galaxies for which the bulge, bar and disk decomposition's were given in literature \citep{Salo.2015}. We have finally used 63 barred galaxies for the study, due to the constraints of matching the simulation models to observed galaxies and the availability of rotation curves. We find that $\approx$92\% of all the barred galaxies in our sample follow the theoretical bar formation criterion, and the outliers are either interacting with companions or have patchy rotation curves. We also find that our criterion is a necessary but not a sufficient condition for bar formation as the effect of disk velocity dispersion cannot be fully accounted for in this criterion as it uses isolated galaxy simulations.

In future, this study can be improved by using precise rotation velocities at disk scale lengths using Integrated Field Unit (IFU) data, as IFU data can provide the rotation velocities at all radii. This is especially important for galaxies that do not have a strong gas component and for which HI or CO rotation curves are not possible to obtain. A much broader study of the bar formation criterion to disk galaxies by calculating the $FB$ constant will improve our understanding of why bars form in some galaxies but not in all galaxies. 

\section{AKNOWLEDGEMENT} \label{A}
This research has made use of the VizieR catalogue access tool, CDS,
Strasbourg, France (DOI : 10.26093/cds/vizier). The original description of the VizieR service was published in 2000, A\&AS 143, 23. We acknowledge the usage of the HyperLeda database (http://leda.univ-lyon1.fr). This research has also made use of the NASA/IPAC Extragalactic Database (NED), which is operated by the Jet Propulsion Laboratory, California Institute of Technology, under contract with the National Aeronautics and Space Administration. We also thank Christopher Mihos for useful suggestion to add value of bar criterion for Milky Way in this study.  Finally, we thank referee for useful comments which has improved the content of this article.

\bibliographystyle{aa} 
\bibliography{Ref.bib}

\begin{thebibliography}{66}
\expandafter\ifx\csname natexlab\endcsname\relax\def\natexlab#1{#1}\fi

\bibitem[{{Aguerri} {et~al.}(2001){Aguerri}, {Balcells}, \&
  {Peletier}}]{Aguerri.2001}
{Aguerri}, J.~A.~L., {Balcells}, M., \& {Peletier}, R.~F. 2001, \aap, 367, 428

\bibitem[{{Aguerri} {et~al.}(2009){Aguerri}, {M{\'e}ndez-Abreu}, \&
  {Corsini}}]{Aguerri.2009}
{Aguerri}, J.~A.~L., {M{\'e}ndez-Abreu}, J., \& {Corsini}, E.~M. 2009, \aap,
  495, 491

\bibitem[{{Athanassoula}(2005)}]{Athanassoula.2005}
{Athanassoula}, E. 2005, \mnras, 358, 1477

\bibitem[{{Athanassoula}(2008)}]{Athanassoula.2008}
{Athanassoula}, E. 2008, \mnras, 390, L69

\bibitem[{{Athanassoula} {et~al.}(2005){Athanassoula}, {Lambert}, \&
  {Dehnen}}]{Athanassoula.2005a}
{Athanassoula}, E., {Lambert}, J.~C., \& {Dehnen}, W. 2005, \mnras, 363, 496

\bibitem[{{Barazza} {et~al.}(2008){Barazza}, {Jogee}, \&
  {Marinova}}]{Barazza.2008}
{Barazza}, F.~D., {Jogee}, S., \& {Marinova}, I. 2008, \apj, 675, 1194

\bibitem[{{Barrera-Ballesteros} {et~al.}(2015){Barrera-Ballesteros},
  {Garc{\'\i}a-Lorenzo}, {Falc{\'o}n-Barroso}, {van de Ven}, {Lyubenova},
  {Wild}, {M{\'e}ndez-Abreu}, {S{\'a}nchez}, {Marquez}, {Masegosa},
  {Monreal-Ibero}, {Ziegler}, {del Olmo}, {Verdes-Montenegro},
  {Garc{\'\i}a-Benito}, {Husemann}, {Mast}, {Kehrig}, {Iglesias-Paramo},
  {Marino}, {Aguerri}, {Walcher}, {V{\'\i}lchez}, {Bomans}, {Cortijo-Ferrero},
  {Gonz{\'a}lez Delgado}, {Bland-Hawthorn}, {McIntosh}, \&
  {Bekerait{\.{e}}}}]{Barrera-Ballesteros.2015}
{Barrera-Ballesteros}, J.~K., {Garc{\'\i}a-Lorenzo}, B., {Falc{\'o}n-Barroso},
  J., {et~al.} 2015, \aap, 582, A21

\bibitem[{{Barway} {et~al.}(2016){Barway}, {Saha}, {Vaghmare}, \&
  {Kembhavi}}]{Barway.2016}
{Barway}, S., {Saha}, K., {Vaghmare}, K., \& {Kembhavi}, A.~K. 2016, \mnras,
  463, L41

\bibitem[{{Baugh} {et~al.}(1996){Baugh}, {Cole}, \& {Frenk}}]{Baugh.1996}
{Baugh}, C.~M., {Cole}, S., \& {Frenk}, C.~S. 1996, \mnras, 283, 1361

\bibitem[{{Beckmann} {et~al.}(2006){Beckmann}, {Gehrels}, {Shrader}, \&
  {Soldi}}]{Beckmann.2006}
{Beckmann}, V., {Gehrels}, N., {Shrader}, C.~R., \& {Soldi}, S. 2006, \apj,
  638, 642

\bibitem[{{Bennett} {et~al.}(2013){Bennett}, {Larson}, {Weiland}, {Jarosik},
  {Hinshaw}, {Odegard}, {Smith}, {Hill}, {Gold}, {Halpern}, {Komatsu}, {Nolta},
  {Page}, {Spergel}, {Wollack}, {Dunkley}, {Kogut}, {Limon}, {Meyer}, {Tucker},
  \& {Wright}}]{Hinshaw.2013}
{Bennett}, C.~L., {Larson}, D., {Weiland}, J.~L., {et~al.} 2013, \apjs, 208, 20

\bibitem[{{Binney} \& {Tremaine}(2008)}]{BT.2008}
{Binney}, J. \& {Tremaine}, S. 2008, {Galactic Dynamics: Second Edition}

\bibitem[{{Bland-Hawthorn} \& {Gerhard}(2016)}]{Bland-Hawthorn.2016}
{Bland-Hawthorn}, J. \& {Gerhard}, O. 2016, \araa, 54, 529

\bibitem[{{Bournaud} {et~al.}(2007){Bournaud}, {Jog}, \&
  {Combes}}]{Bournaud.2007}
{Bournaud}, F., {Jog}, C.~J., \& {Combes}, F. 2007, \aap, 476, 1179

\bibitem[{{Bovy} \& {Rix}(2013)}]{Bovy.2013}
{Bovy}, J. \& {Rix}, H.-W. 2013, \apj, 779, 115

\bibitem[{{Combes} {et~al.}(1990){Combes}, {Debbasch}, {Friedli}, \&
  {Pfenniger}}]{Combes.1990}
{Combes}, F., {Debbasch}, F., {Friedli}, D., \& {Pfenniger}, D. 1990, \aap,
  233, 82

\bibitem[{{Das} {et~al.}(2008){Das}, {Laurikainen}, {Salo}, \&
  {Buta}}]{Das.2008}
{Das}, M., {Laurikainen}, E., {Salo}, H., \& {Buta}, R. 2008, \apss, 317, 163

\bibitem[{{Das} {et~al.}(2003){Das}, {Teuben}, {Vogel}, {Regan}, {Sheth},
  {Harris}, \& {Jefferys}}]{Das.2003}
{Das}, M., {Teuben}, P.~J., {Vogel}, S.~N., {et~al.} 2003, \apj, 582, 190

\bibitem[{{Debattista} {et~al.}(2006){Debattista}, {Mayer}, {Carollo}, {Moore},
  {Wadsley}, \& {Quinn}}]{Debattista.2006}
{Debattista}, V.~P., {Mayer}, L., {Carollo}, C.~M., {et~al.} 2006, \apj, 645,
  209

\bibitem[{{D{\'\i}az-Garc{\'\i}a} {et~al.}(2016){D{\'\i}az-Garc{\'\i}a},
  {Salo}, \& {Laurikainen}}]{Diaz-Garcia.2016}
{D{\'\i}az-Garc{\'\i}a}, S., {Salo}, H., \& {Laurikainen}, E. 2016, \aap, 596,
  A84

\bibitem[{{Efstathiou} {et~al.}(1982){Efstathiou}, {Lake}, \&
  {Negroponte}}]{Efstathiou.1982}
{Efstathiou}, G., {Lake}, G., \& {Negroponte}, J. 1982, \mnras, 199, 1069

\bibitem[{{Eggen} {et~al.}(1962){Eggen}, {Lynden-Bell}, \&
  {Sandage}}]{Eggen.1962}
{Eggen}, O.~J., {Lynden-Bell}, D., \& {Sandage}, A.~R. 1962, \apj, 136, 748

\bibitem[{{Elmegreen} {et~al.}(1995){Elmegreen}, {Sundin}, {Kaufman}, {Brinks},
  \& {Elmegreen}}]{Elmegreen.1995}
{Elmegreen}, B.~G., {Sundin}, M., {Kaufman}, M., {Brinks}, E., \& {Elmegreen},
  D.~M. 1995, \apj, 453, 139

\bibitem[{{Erwin} {et~al.}(2015){Erwin}, {Saglia}, {Fabricius}, {Thomas},
  {Nowak}, {Rusli}, {Bender}, {Vega Beltr{\'a}n}, \& {Beckman}}]{Erwin.2015}
{Erwin}, P., {Saglia}, R.~P., {Fabricius}, M., {et~al.} 2015, \mnras, 446, 4039

\bibitem[{{Fisher} \& {Drory}(2008)}]{FD.2008}
{Fisher}, D.~B. \& {Drory}, N. 2008, \aj, 136, 773

\bibitem[{{Gerhard}(2002)}]{Gerhard.2002}
{Gerhard}, O. 2002, Astronomical Society of the Pacific Conference Series, Vol.
  273, {The Galactic Bar}, ed. G.~S. {Da Costa}, E.~M. {Sadler}, \&
  H.~{Jerjen}, 73

\bibitem[{{Graham} \& {Worley}(2008)}]{GW.2008}
{Graham}, A.~W. \& {Worley}, C.~C. 2008, \mnras, 388, 1708

\bibitem[{{Guo} {et~al.}(2020){Guo}, {Du}, {Ho}, {Debattista}, \&
  {Zhao}}]{Guo.2020}
{Guo}, M., {Du}, M., {Ho}, L.~C., {Debattista}, V.~P., \& {Zhao}, D. 2020,
  \apj, 888, 65

\bibitem[{{Heller} \& {Shlosman}(1994)}]{Heller.1994}
{Heller}, C.~H. \& {Shlosman}, I. 1994, \apj, 424, 84

\bibitem[{{Hohl}(1976)}]{Hohl.1976}
{Hohl}, F. 1976, \aj, 81, 30

\bibitem[{{Hopkins} {et~al.}(2010){Hopkins}, {Bundy}, {Croton}, {Hernquist},
  {Keres}, {Khochfar}, {Stewart}, {Wetzel}, \& {Younger}}]{Hopkins.2010}
{Hopkins}, P.~F., {Bundy}, K., {Croton}, D., {et~al.} 2010, \apj, 715, 202

\bibitem[{{Hopkins} {et~al.}(2009){Hopkins}, {Cox}, {Younger}, \&
  {Hernquist}}]{Hopkins.2009}
{Hopkins}, P.~F., {Cox}, T.~J., {Younger}, J.~D., \& {Hernquist}, L. 2009,
  \apj, 691, 1168

\bibitem[{{Kataria} \& {Das}(2018)}]{KD.2018}
{Kataria}, S.~K. \& {Das}, M. 2018, \mnras, 475, 1653

\bibitem[{{Kataria} \& {Das}(2019)}]{Kataria.2019}
{Kataria}, S.~K. \& {Das}, M. 2019, \apj, 886, 43

\bibitem[{{Kauffmann} {et~al.}(1993){Kauffmann}, {White}, \&
  {Guiderdoni}}]{Kauffmann.1993}
{Kauffmann}, G., {White}, S.~D.~M., \& {Guiderdoni}, B. 1993, \mnras, 264, 201

\bibitem[{{Kormendy} \& {Kennicutt}(2004)}]{Kormendy.2004}
{Kormendy}, J. \& {Kennicutt}, Jr., R.~C. 2004, \araa, 42, 603

\bibitem[{{Laurikainen} {et~al.}(2007){Laurikainen}, {Salo}, {Buta}, \&
  {Knapen}}]{Laurikainen.2007}
{Laurikainen}, E., {Salo}, H., {Buta}, R., \& {Knapen}, J.~H. 2007, \mnras,
  381, 401

\bibitem[{{Lelli} {et~al.}(2016){Lelli}, {McGaugh}, \&
  {Schombert}}]{Lelli.2016}
{Lelli}, F., {McGaugh}, S.~S., \& {Schombert}, J.~M. 2016, \aj, 152, 157

\bibitem[{{Licquia} {et~al.}(2016){Licquia}, {Newman}, \&
  {Bershady}}]{Licquia.2016}
{Licquia}, T.~C., {Newman}, J.~A., \& {Bershady}, M.~A. 2016, \apj, 833, 220

\bibitem[{{Makarov} {et~al.}(2014){Makarov}, {Prugniel}, {Terekhova},
  {Courtois}, \& {Vauglin}}]{Makarov.2014}
{Makarov}, D., {Prugniel}, P., {Terekhova}, N., {Courtois}, H., \& {Vauglin},
  I. 2014, \aap, 570, A13

\bibitem[{{Martinez-Valpuesta} {et~al.}(2006){Martinez-Valpuesta}, {Shlosman},
  \& {Heller}}]{Martinez.2006}
{Martinez-Valpuesta}, I., {Shlosman}, I., \& {Heller}, C. 2006, \apj, 637, 214

\bibitem[{{McGaugh}(2018)}]{Mcgaugh.2018}
{McGaugh}, S.~S. 2018, Research Notes of the American Astronomical Society, 2,
  156

\bibitem[{{McMillan}(2011)}]{McMillan.2011}
{McMillan}, P.~J. 2011, \mnras, 414, 2446

\bibitem[{{Naab} {et~al.}(2014){Naab}, {Oser}, {Emsellem}, {Cappellari},
  {Krajnovi{\'c}}, {McDermid}, {Alatalo}, {Bayet}, {Blitz}, {Bois}, {Bournaud},
  {Bureau}, {Crocker}, {Davies}, {Davis}, {de Zeeuw}, {Duc}, {Hirschmann},
  {Johansson}, {Khochfar}, {Kuntschner}, {Morganti}, {Oosterloo}, {Sarzi},
  {Scott}, {Serra}, {Ven}, {Weijmans}, \& {Young}}]{Naab.2014}
{Naab}, T., {Oser}, L., {Emsellem}, E., {et~al.} 2014, \mnras, 444, 3357

\bibitem[{{Norman} {et~al.}(1996){Norman}, {Sellwood}, \&
  {Hasan}}]{Norman.1996}
{Norman}, C.~A., {Sellwood}, J.~A., \& {Hasan}, H. 1996, \apj, 462, 114

\bibitem[{{Ostriker} \& {Peebles}(1973)}]{OP.1973}
{Ostriker}, J.~P. \& {Peebles}, P.~J.~E. 1973, \apj, 186, 467

\bibitem[{{Paturel} {et~al.}(2003){Paturel}, {Theureau}, {Bottinelli},
  {Gouguenheim}, {Coudreau-Durand}, {Hallet}, \& {Petit}}]{Paturel.2003}
{Paturel}, G., {Theureau}, G., {Bottinelli}, L., {et~al.} 2003, \aap, 412, 57

\bibitem[{{Peng} {et~al.}(2002){Peng}, {Ho}, {Impey}, \& {Rix}}]{Peng.2002}
{Peng}, C.~Y., {Ho}, L.~C., {Impey}, C.~D., \& {Rix}, H.-W. 2002, \aj, 124, 266

\bibitem[{{Peng} {et~al.}(2010){Peng}, {Ho}, {Impey}, \& {Rix}}]{Peng.2010}
{Peng}, C.~Y., {Ho}, L.~C., {Impey}, C.~D., \& {Rix}, H.-W. 2010, \aj, 139,
  2097

\bibitem[{{Pfenniger} \& {Norman}(1990)}]{Pfenniger.1990}
{Pfenniger}, D. \& {Norman}, C. 1990, \apj, 363, 391

\bibitem[{{Portail} {et~al.}(2015){Portail}, {Wegg}, {Gerhard}, \&
  {Martinez-Valpuesta}}]{Portail.2015}
{Portail}, M., {Wegg}, C., {Gerhard}, O., \& {Martinez-Valpuesta}, I. 2015,
  \mnras, 448, 713

\bibitem[{{Qu} {et~al.}(2011){Qu}, {Di Matteo}, {Lehnert}, {van Driel}, \&
  {Jog}}]{Qu.2011}
{Qu}, Y., {Di Matteo}, P., {Lehnert}, M.~D., {van Driel}, W., \& {Jog}, C.~J.
  2011, \aap, 535, A5

\bibitem[{{Raha} {et~al.}(1991){Raha}, {Sellwood}, {James}, \&
  {Kahn}}]{Raha.1991}
{Raha}, N., {Sellwood}, J.~A., {James}, R.~A., \& {Kahn}, F.~D. 1991, \nat,
  352, 411

\bibitem[{{Regan} \& {Teuben}(2004)}]{Regan.2004}
{Regan}, M.~W. \& {Teuben}, P.~J. 2004, \apj, 600, 595

\bibitem[{{Salo} {et~al.}(2015){Salo}, {Laurikainen}, {Laine}, {Comer{\'o}n},
  {Gadotti}, {Buta}, {Sheth}, {Zaritsky}, {Ho}, {Knapen}, {Athanassoula},
  {Bosma}, {Laine}, {Cisternas}, {Kim}, {Mu{\~n}oz-Mateos}, {Regan}, {Hinz},
  {Gil de Paz}, {Menendez-Delmestre}, {Mizusawa}, {Erroz-Ferrer}, {Meidt}, \&
  {Querejeta}}]{Salo.2015}
{Salo}, H., {Laurikainen}, E., {Laine}, J., {et~al.} 2015, \apjs, 219, 4

\bibitem[{{Sellwood}(1980)}]{Sellwood.1980}
{Sellwood}, J.~A. 1980, \aap, 89, 296

\bibitem[{{Shen} \& {Sellwood}(2004)}]{Shen.2004}
{Shen}, J. \& {Sellwood}, J.~A. 2004, \apj, 604, 614

\bibitem[{{Sheth} {et~al.}(2008){Sheth}, {Elmegreen}, {Elmegreen}, {Capak},
  {Abraham}, {Athanassoula}, {Ellis}, {Mobasher}, {Salvato}, {Schinnerer},
  {Scoville}, {Spalsbury}, {Strubbe}, {Carollo}, {Rich}, \&
  {West}}]{Sheth.2008}
{Sheth}, K., {Elmegreen}, D.~M., {Elmegreen}, B.~G., {et~al.} 2008, \apj, 675,
  1141

\bibitem[{{Skibba} {et~al.}(2012){Skibba}, {Masters}, {Nichol}, {Zehavi},
  {Hoyle}, {Edmondson}, {Bamford}, {Cardamone}, {Keel}, {Lintott}, \&
  {Schawinski}}]{Skibba.2012}
{Skibba}, R.~A., {Masters}, K.~L., {Nichol}, R.~C., {et~al.} 2012, \mnras, 423,
  1485

\bibitem[{{Sofue} {et~al.}(1999){Sofue}, {Tutui}, {Honma}, {Tomita},
  {Takamiya}, {Koda}, \& {Takeda}}]{Sofue.1999}
{Sofue}, Y., {Tutui}, Y., {Honma}, M., {et~al.} 1999, \apj, 523, 136

\bibitem[{{Telles} \& {Terlevich}(1995)}]{Telles.1995}
{Telles}, E. \& {Terlevich}, R. 1995, \mnras, 275, 1

\bibitem[{{Thomas} {et~al.}(2002){Thomas}, {Dunne}, {Clemens}, {Alexand er},
  {Eales}, \& {Green}}]{Thomas.2002}
{Thomas}, H.~C., {Dunne}, L., {Clemens}, M.~S., {et~al.} 2002, \mnras, 329, 747

\bibitem[{{Toomre}(1981)}]{Toomre.1981}
{Toomre}, A. 1981, in Structure and Evolution of Normal Galaxies, ed. S.~M.
  {Fall} \& D.~{Lynden-Bell}, 111--136

\bibitem[{{Valluri}(1993)}]{Monica.1993}
{Valluri}, M. 1993, \apj, 408, 57

\bibitem[{{Watkins} {et~al.}(2014){Watkins}, {Mihos}, {Harding}, \&
  {Feldmeier}}]{Watkins.2014}
{Watkins}, A.~E., {Mihos}, J.~C., {Harding}, P., \& {Feldmeier}, J.~J. 2014,
  \apj, 791, 38

\bibitem[{{Yurin} \& {Springel}(2014)}]{Yurin.2014}
{Yurin}, D. \& {Springel}, V. 2014, \mnras, 444, 62

\end{thebibliography}

\longtab{
\begin{longtable}{lrrrrrrl}
\caption{Galaxy Samples used for Bar formation criterion Study}\\
\hline
\hline
Galaxy Name	&	Morphology	&	$V_{rot}$	&   Bulge magnitude	& B/T & Inclination  &	$FB$  \\
  &  & (Km/s) &(3.6 micron AB magnitude) &  & (degrees) & \\
\hline
\endfirsthead
\caption{Continued.} \\
\hline
\hline 
Galaxy Name	&	Morphology	&	$V_{rot}$	&	Bulge magnitude	& B/T & Inclination &	$FB$  \\
 
  &  & (Km/s) &       (3.6 micron AB magnitude) &  & (degrees) & \\
\hline
\endhead
\hline
\endfoot
\hline
\endlastfoot

ESO\,027-001	&	SBc	&	87.2	&	13.985	&	0.087	&	37	&	0.019	\\
ESO\,509-026	&	SABm	&	21.8	&	15.55	&	0.232	&	65.5	&	0.181	\\
IC\,4214	&	SBa	&	204.3	&	11.917	&	0.299	&	55.4	&	0.022	\\
NGC\,0210	&	SABb	&	154.2	&	11.965	&	0.291	&	55.4	&	0.009	\\
NGC\,0254	&	S0-a	&	148.7	&	12.428	&	0.333	&	57.2	&	0.016	\\
NGC\,0615	&	Sb	&	183.4	&	12.819	&	0.201	&	74	&	0.012	\\
NGC\,0936	&	S0-a	&	341.3	&	11.275	&	0.195	&	50.4	&	0.004	\\
NGC\,1015	&	Sa	&	120.7	&	13.508	&	0.209	&	35.5	&	0.017	\\
NGC\,1022	&	SBa	&	86.5	&	12.269	&	0.278	&	59.9	&	0.064	\\
NGC\,1232	&	SABc	&	185.2	&	13.814	&	0.022	&	32.7	&	0.001	\\
NGC\,1326	&	S0-a	&	138.4	&	11.224	&	0.318	&	52.7	&	0.029	\\
NGC\,1367	&	Sa	&	232	&	12.382	&	0.138	&	51.9	&	0.008	\\
NGC\,1415	&	S0-a	&	163.6	&	12.598	&	0.174	&	77.7	&	0.002	\\
NGC\,1533	&	E-S0	&	132.7	&	11.697	&	0.249	&	64.8	&	0.083	\\
NGC\,2633	&	Sb	&	142.6	&	12.287	&	0.366	&	53.8	&	0.03	\\
NGC\,2750	&	SABc	&	86.4	&	13.884	&	0.148	&	45.4	&	0.007	\\
NGC\,2893	&	S0-a	&	107.5	&	13.881	&	0.311	&	39.1	&	0.033	\\
NGC\,2962	&	S0-a	&	203.9	&	12.722	&	0.238	&	72.3	&	0.014	\\
NGC\,2968	&	Sa	&	122.9	&	11.905	&	0.363	&	53.3	&	0.044	\\
NGC\,3061	&	Sc	&	141.4	&	17.023	&	0.022	&	39.6	&	0.001	\\
NGC\,3166	&	S0-a	&	103.9	&	11.123	&	0.274	&	56.2	&	0.083	\\
NGC\,3254	&	Sbc	&	207.8	&	13.398	&	0.158	&	64.4	&	0.009	\\
NGC\,3266	&	S0	&	264.5	&	13.811	&	0.248	&	34.7	&	0.005	\\
NGC\,3424	&	SBb	&	167.9	&	13.19	&	0.178	&	79.2	&	0.011	\\
NGC\,3489	&	S0-a	&	54.4	&	11.2	&	0.321	&	63.7	&	0.377	\\
NGC\,3507	&	SBb	&	117.9	&	13.754	&	0.068	&	31.9	&	0.006	\\
NGC\,3626	&	S0-a	&	190.4	&	11.939	&	0.297	&	55.8	&	0.018	\\
NGC\,3627	&	Sb	&	174.7	&	10.882	&	0.091	&	67.5	&	0.017	\\
NGC\,3684	&	Sbc	&	120.2	&	15.328	&	0.034	&	50.7	&	0.004	\\
NGC\,3686	&	SBbc	&	129.7	&	14.685	&	0.029	&	41.7	&	0.002	\\
NGC\,3705	&	SABa	&	167.9	&	13.012	&	0.091	&	72.2	&	0.016	\\
NGC\,3930	&	Sc	&	96.9	&	15.853	&	0.034	&	43	&	0.001	\\
NGC\,3953	&	Sbc	&	215.3	&	12.406	&	0.064	&	62.3	&	0.002	\\
NGC\,4045	&	Sa	&	169.1	&	12.586	&	0.245	&	55.9	&	0.019	\\
NGC\,4102	&	SABb	&	158	&	11.592	&	0.261	&	58.7	&	0.044	\\
NGC\,4133	&	SABb	&	166.1	&	13.904	&	0.147	&	51.3	&	0.01	\\
NGC\,4267	&	E-S0	&	60.2	&	11.638	&	0.29	&	57.4	&	0.174	\\
NGC\,4319	&	SBab	&	112.5	&	13.758	&	0.099	&	72.5	&	0.013	\\
NGC\,4355	&	SABa	&	47	&	13.83	&	0.316	&	68.1	&	0.16	\\
NGC\,4569	&	Sab	&	185.8	&	11.222	&	0.146	&	70.9	&	0.023	\\
NGC\,4606	&	SBa	&	70.2	&	13.608	&	0.17	&	62.7	&	0.009	\\
NGC\,4795	&	SBa	&	191.8	&	13.533	&	0.149	&	53.6	&	0.01	\\
NGC\,4897	&	Sbc	&	168.1	&	13.726	&	0.2	&	38.3	&	0.011	\\
NGC\,4984	&	S0-a	&	125.3	&	11.092	&	0.382	&	47.1	&	0.041	\\
NGC\,5005	&	SABb	&	251	&	11	&	0.152	&	77.1	&	0.023	\\
NGC\,5195	&	SBa	&	120.3	&	10.122	&	0.253	&	40.5	&	0.024	\\
NGC\,5205	&	SBbc	&	136.5	&	13.734	&	0.287	&	56	&	0.015	\\
NGC\,5377	&	Sa	&	177.1	&	11.732	&	0.394	&	77.2	&	0.017	\\
NGC\,5443	&	Sb	&	167.6	&	13.608	&	0.144	&	77.4	&	0.007	\\
NGC\,5448	&	Sa	&	208	&	12.517	&	0.282	&	64.4	&	0.009	\\
NGC\,5473	&	E-S0	&	158	&	12.299	&	0.276	&	47.7	&	0.043	\\
NGC\,5566	&	SBab	&	204	&	11.511	&	0.228	&	75.6	&	0.01	\\
NGC\,5713	&	SABb	&	107.9	&	13.571	&	0.069	&	48.2	&	0.025	\\
NGC\,5728	&	Sa	&	224.4	&	11.934	&	0.263	&	53.1	&	0.008	\\
NGC\,5878	&	Sb	&	214.5	&	12.795	&	0.148	&	73.8	&	0.007	\\
NGC\,6278	&	S0	&	76	&	13.007	&	0.277	&	78.8	&	0.151	\\
NGC\,6923	&	SBb	&	191.1	&	13.334	&	0.137	&	64.2	&	0.008	\\
NGC\,7280	&	S0-a	&	129.4	&	13.23	&	0.223	&	58.7	&	0.012	\\
NGC\,7465	&	S0	&	95	&	13.092	&	0.357	&	64.2	&	0.113	\\
NGC\,7479	&	SBbc	&	238.3	&	12.958	&	0.083	&	43	&	0.004	\\
NGC\,7531	&	SABb	&	161.2	&	13.115	&	0.138	&	68.9	&	0.011	\\
NGC\,7582	&	SBab	&	194.7	&	11.185	&	0.252	&	68	&	0.016	\\
NGC\,7731	&	SBa	&	92.7	&	14.839	&	0.203	&	45.5	&	0.019	\\

\end{longtable}
\noindent
Note: Galaxy Morphology, $V_{rot}$ and inclination angle is taken from Hyperleda database. Bulge magnitude as well as \\ Bulge to total ratio is taken from \citet{Salo.2015}. }

\end{document}